\documentclass[11pt,a4paper]{article}
\usepackage[margin=26mm]{geometry}
\usepackage[T1]{fontenc}
\usepackage{lmodern}
\usepackage{cmap}
\input{glyphtounicode}
\usepackage{amsmath,amssymb,amsthm,bm,booktabs,array}
\usepackage{graphicx}
\usepackage{microtype}
\usepackage{float}
\usepackage[hidelinks]{hyperref}
\hypersetup{pdftitle={The Geometry of Predictive Precision: Optimal Allocation and Attainable Gain},pdfauthor={Song-Ju Kim}}
\usepackage{caption}
\graphicspath{{figures/}}
\newtheorem{proposition}{Proposition}

\newcommand{\tr}{\operatorname{tr}}
\newcommand{\diag}{\operatorname{diag}}

\title{The Geometry of Predictive Precision: Optimal Allocation and Attainable Gain}
\author{Song-Ju Kim\\[3pt]\small SOBIN Institute LLC, Kawanishi, Hyogo, Japan\\[-1pt]\small \texttt{kim@sobin.org}}
\date{}
\begin{document}
\maketitle
\begin{abstract}
A finite-time prediction can amplify small initial errors very differently across state-space directions. Which directions deserve precision depends on both the dynamics and the terminal quantity of interest. We minimize task-weighted terminal error over local covariance shapes of fixed generalized volume. The unique optimum is the inverse of a predictive metric obtained by pulling the terminal requirement through the tangent map; it equalizes propagated uncertainty in the required terminal metric. A spectral expression gives the gain over uniform allocation. We then distinguish two constraints with different physical consequences: limited anisotropy caps the gain, while coordinate-diagonal allocation can lose almost all the benefit if sensitive directions are rotated away from accessible axes. Under dominant finite-time stretching and a fixed anisotropy cap, optimal allocations for different terminal requirements also approach a common shape. Standard-map and double-pendulum examples, together with a direct finite-amplitude check, illustrate these effects. The resulting criteria guide task-aware sensing, state encoding, and precision-controlled simulation by identifying when accessible directions matter as much as nominal resolution.
\end{abstract}


\section{Introduction}\label{sec:intro}

Consider predicting a mechanical system for a specified time interval. A small error in one initial direction may be amplified strongly, while an equally sized error in another direction may contract. Moreover, accurately predicting one terminal angle is not the same task as accurately predicting another. When local state precision is limited, treating every initial direction equally can therefore spend resolution on distinctions that barely matter to the required prediction.

This paper asks a concrete question: \emph{given a fixed volume of local uncertainty, how should its directional shape be chosen to minimize a specified terminal prediction error?} The essential distinction is between two choices that are often mixed together: the dynamics determines how an initial error propagates, while the \emph{terminal requirement} determines which propagated errors count. The present representation should respond to both. Figure~\ref{fig:principle} previews the answer: initial uncertainty is shaped against the finite-time stretching so that the final error becomes balanced in the metric selected by the prediction task.

Finite-time tangent maps and Cauchy--Green tensors characterize deformation along trajectories \cite{Haller2011,Boffetta2002}. Singular-vector forecasting has long recognized that the terminal norm changes which initial directions are important \cite{Palmer1998}. Determinant--trace optimization and continuous precision allocation provide well-established matrix tools \cite{Pukelsheim2006,GrayNeuhoff1998}; task-aware quantization similarly distinguishes prediction-relevant error from uniform reconstruction error \cite{Shlezinger2019}. The contribution here is an explicit finite-horizon allocation problem in which these elements are put together: a prescribed terminal requirement induces a predictive metric, a fixed-volume covariance class admits a unique optimal shape, and the resulting gain can be compared directly with the geometric distortion required to realize it.

The argument proceeds from the required prediction to the geometry of present uncertainty: Section~\ref{sec:setting} defines the task and volume constraint; Section~\ref{sec:matching} obtains the optimal local allocation; Section~\ref{sec:limits} asks how finite directional range and accessible coordinates affect its benefit; and Section~\ref{sec:examples} tests these consequences in nonlinear Hamiltonian systems. Section~\ref{sec:meaning} connects the benchmark to its physical implementation and finite-amplitude validity.

\section{Prediction task, local uncertainty, and physical meaning}\label{sec:setting}

\subsection{From a terminal requirement to a present sensitivity metric}

Let $\bm{x}_t\in\mathbb R^n$ denote the state at time $t$, with $n$ the number of state variables. Its finite-time evolution to $T>t$ is $\bm{x}_T=\varphi_{t,T}(\bm{x}_t)$. For a continuous system $\dot{\bm{x}}=\bm f(\bm{x})$, $\varphi$ is the flow; for an iterated map, it is the corresponding composition. The tangent map along a reference trajectory is
\begin{equation}
 \Phi_{t,T}=D\varphi_{t,T}(\bm{x}_t),\qquad
 \delta\bm{x}_T\simeq\Phi_{t,T}\delta\bm{x}_t.
 \label{eq:phi}
\end{equation}
Here $\delta\bm{x}_t$ is a small perturbation to the present state and $\delta\bm{x}_T$ is its leading-order terminal effect. The map is assumed nonsingular on the interval under study.

The prediction task is represented by a symmetric positive-definite matrix $Q_R$. It assigns the terminal quadratic loss
\begin{equation}
 \ell_R(\delta\bm{x}_T)=\delta\bm{x}_T^{\mathsf T}Q_R\delta\bm{x}_T.
 \label{eq:loss}
\end{equation}
For example, after fixing physically meaningful reference scales, $Q_R=I$ weights every terminal coordinate equally; a larger weight on one angle means that angle matters more to the prediction. Pulling this loss back to the present gives
\begin{equation}
 \ell_R\simeq\delta\bm{x}_t^{\mathsf T}A_{t,T}\delta\bm{x}_t,
 \qquad\boxed{A_{t,T}=\Phi_{t,T}^{\mathsf T}Q_R\Phi_{t,T}}.
 \label{eq:A}
\end{equation}
Thus $A_{t,T}$ measures \emph{present errors by their future consequences}. We refer to it as the task-weighted predictive metric. The standard pullback supplies a concrete optimization objective for allocating local precision to a specified terminal prediction.

\begin{table}[H]
\centering
\caption{Core quantities and their interpretation. The reference covariance $\Sigma_0$ fixes coordinate scales, whereas the terminal metric $Q_R$ fixes the prediction task.}
\label{tab:notation}
\small
\begin{tabular}{@{}p{0.18\linewidth}p{0.74\linewidth}@{}}
\toprule
Symbol & Meaning \\
\midrule
$n$, $t$, $T$ & State dimension, present time, and terminal prediction time. \\
$\Phi_{t,T}$ & Tangent map: the linear response of the terminal state to a present perturbation. \\
$Q_R$, $A_{t,T}$ & Terminal error metric and its pullback to the present state space. \\
$\Sigma$, $\Sigma_0$ & Adjustable present error covariance and fixed reference covariance defining the coordinate scales. \\
$\rho$ & Common variance scale: $\det(\Sigma_0^{-1}\Sigma)=\rho^n$. \\
$C$, $Y$ & Reference-scaled predictive metric and adjustable covariance, respectively; see Eq.~\eqref{eq:CY}. \\
$G$, $\kappa_{\max}$ & Ratio of reference-isotropic to optimized loss, and the permitted condition number of the reference-scaled covariance $Y$. \\
\bottomrule
\end{tabular}
\end{table}

The distinction between dynamics and task is operationally important. Even if $\Phi_{t,T}$ stays unchanged, modifying $Q_R$ changes $A_{t,T}$ and therefore changes which present errors are costly. Table~\ref{tab:notation} summarizes the notation used throughout the paper.

\subsection{What is held fixed when precision is reallocated?}

We describe present local representation error as a centered perturbation with positive-definite covariance $\Sigma$. Under the tangent approximation, its expected terminal loss is
\begin{equation}
 \mathcal L_R(\Sigma)=\mathbb E[\ell_R]=\tr(A_{t,T}\Sigma).
 \label{eq:L}
\end{equation}
The covariance is an abstract local error model; its eigenvectors indicate principal error directions and its eigenvalues specify the associated variances. No Gaussian assumption is needed for Eq.~\eqref{eq:L}.

A fixed positive-definite reference covariance $\Sigma_0$ supplies the physical scaling of the state coordinates. We compare covariance shapes at a shared generalized uncertainty volume by imposing
\begin{equation}
 \boxed{\det(\Sigma_0^{-1}\Sigma)=\rho^n,\qquad\rho>0.}
 \label{eq:volume}
\end{equation}
The volume of a covariance ellipsoid at fixed Mahalanobis radius is proportional to $\sqrt{\det\Sigma}$; Eq.~\eqref{eq:volume} therefore holds its volume relative to $\Sigma_0$ fixed. The scalar $\rho$ is the variance multiplier of the reference-isotropic choice $\Sigma_{\mathrm{iso}}=\rho\Sigma_0$. We optimize the \emph{shape} at the same volume, not the total amount of uncertainty.

For clarity, define the dimensionless, reference-scaled matrices
\begin{equation}
 Y=\Sigma_0^{-1/2}\Sigma\Sigma_0^{-1/2},\qquad
 C=\Sigma_0^{1/2}A_{t,T}\Sigma_0^{1/2}.
 \label{eq:CY}
\end{equation}
Then $\det Y=\rho^n$ and $\mathcal L_R=\tr(CY)$. The matrix $C$ combines the terminal task, the finite-time dynamics, and the calibration of the present coordinates. Its eigenvalues describe the relative terminal sensitivity of unit-scaled present errors.

\section{The geometry-matching solution}\label{sec:matching}

\begin{proposition}[Optimal local precision allocation]\label{prop:opt}
Suppose $Q_R>0$, $\Sigma_0>0$, and $\Phi_{t,T}$ is nonsingular. Among all $\Sigma>0$ satisfying Eq.~\eqref{eq:volume}, the expected terminal loss in Eq.~\eqref{eq:L} has the unique minimizer
\begin{equation}
 \boxed{\Sigma^*=c\,A_{t,T}^{-1},\quad
 c=\rho\,(\det C)^{1/n},}
 \label{eq:sigmaopt}
\end{equation}
with minimum
\begin{equation}
 \boxed{\mathcal L_{\min}=n\rho(\det C)^{1/n}.}
 \label{eq:Lmin}
\end{equation}
\end{proposition}

\begin{proof}
In the variables of Eq.~\eqref{eq:CY}, minimize $\tr(CY)$ with $\det Y=\rho^n$. The positive matrix $Z=C^{1/2}YC^{1/2}$ satisfies $\det Z=\rho^n\det C$. The arithmetic--geometric mean inequality gives
\begin{equation}
 \tr Z\ge n(\det Z)^{1/n}=n\rho(\det C)^{1/n},
\end{equation}
with equality exactly when $Z$ is proportional to $I$. Hence $Y^*=\rho(\det C)^{1/n}C^{-1}$, which transforms back to Eq.~\eqref{eq:sigmaopt}. The equality condition also gives uniqueness.
\end{proof}

\begin{figure}[t]
\centering
\includegraphics[width=0.90\linewidth]{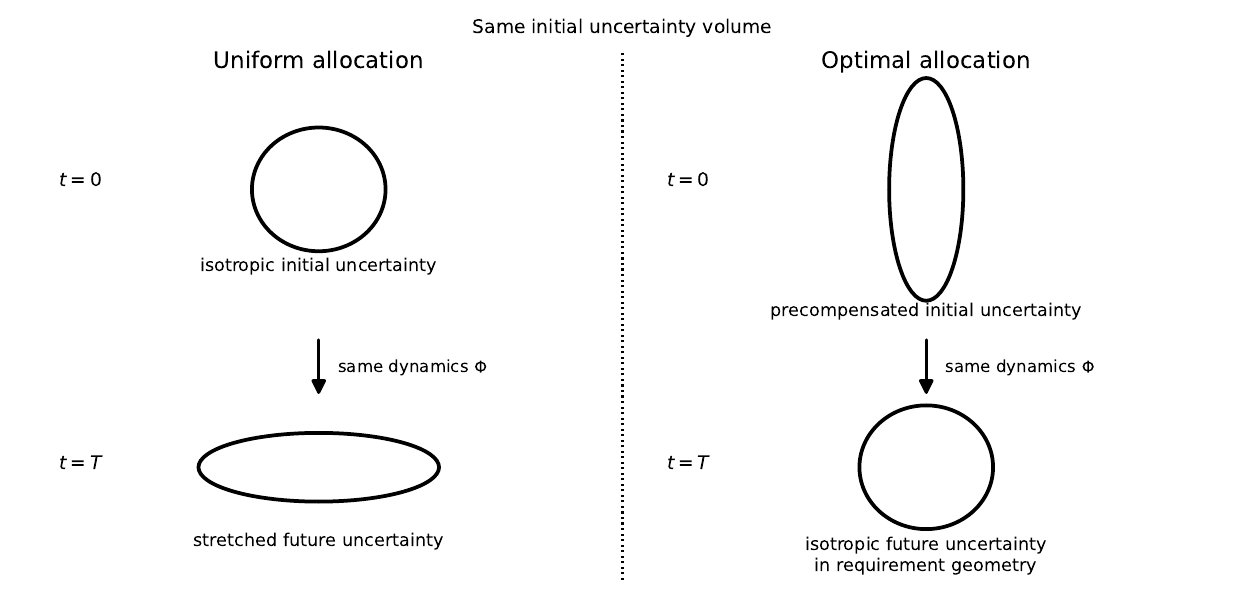}
\caption{Why nonuniform present precision can improve terminal prediction. Both initial ellipses have equal area and experience the same area-preserving deformation. Uniform initial uncertainty (left) becomes stretched at the terminal time. The task-matched initial ellipse (right) anticipates the stretch and produces a round terminal error ellipse in normalized task coordinates. The schematic illustrates the tangent-model isotropization in Eq.~\eqref{eq:isotropize}.}
\label{fig:principle}
\end{figure}

The result has a simple physical interpretation. A direction that magnifies terminal error strongly should receive a smaller present error variance; a direction that matters little can receive a larger variance. The optimum apportions uncertainty so that no terminal direction remains disproportionately costly. Formally, the requirement-normalized terminal covariance is
\begin{equation}
 W_T=Q_R^{1/2}\Phi_{t,T}\Sigma\Phi_{t,T}^{\mathsf T}Q_R^{1/2}.
 \label{eq:W}
\end{equation}
At the optimum,
\begin{equation}
 \boxed{W_T(\Sigma^*)=cI.}
 \label{eq:isotropize}
\end{equation}
The optimal initial representation is generally anisotropic, but its propagated error is isotropic \emph{as judged by the terminal task}. This is the terminal-isotropization principle illustrated in Fig.~\ref{fig:principle}.

\subsection{How much does matching improve prediction?}

The reference-isotropic choice has loss $\mathcal L_{\mathrm{iso}}=\rho\tr C$. Dividing by Eq.~\eqref{eq:Lmin} gives the ideal relative gain
\begin{equation}
 \boxed{G_R=\frac{\mathcal L_{\mathrm{iso}}}{\mathcal L_{\min}}=
 \frac{\tr C}{n(\det C)^{1/n}}\ge 1.}
 \label{eq:gain}
\end{equation}
This formula says that the potential improvement depends on the \emph{spread of directional sensitivities}, not on the overall covariance volume. If all eigenvalues of $C$ are equal, uniform allocation is already optimal and $G_R=1$. If they differ, matching the allocation reduces the weighted terminal loss. Scaling $Q_R$ uniformly changes the absolute loss but not the gain or optimal shape.

The ratio in Eq.~\eqref{eq:gain} is invariant under a consistent invertible change of state coordinates, since both $A_{t,T}$ and the reference covariance transform with those coordinates (Appendix~\ref{app:coordinates}). This matters when combining quantities such as angles and canonical momenta: the conclusion should not depend on a passive change of units.

\section{From optimal geometry to attainable prediction}\label{sec:limits}

\subsection{Finite-time stretching and required anisotropy}

In normalized coordinates with $Q_R=\Sigma_0=I$, the predictive metric is the finite-time Cauchy--Green tensor $C=\Phi^{\mathsf T}\Phi$. If its singular values are $s_1,\ldots,s_n>0$, Eq.~\eqref{eq:gain} becomes
\begin{equation}
 G=\frac{\sum_i s_i^2}{n(\prod_i s_i^2)^{1/n}}.
 \label{eq:spectrum}
\end{equation}
For an area-preserving two-dimensional system, the singular values are $s$ and $s^{-1}$ ($s\ge1$). The ideal gain and the condition number of the optimal covariance are then
\begin{equation}
 \boxed{G=\tfrac12(s^2+s^{-2}),\qquad
 \kappa(\Sigma^*)=s^4.}
 \label{eq:Gandk}
\end{equation}
Here $\kappa$ denotes the ratio of the largest to smallest covariance eigenvalue in the normalized coordinates. Eliminating $s$ gives
\begin{equation}
 \boxed{G=\frac12\left(\sqrt{\kappa(\Sigma^*)}+\frac{1}{\sqrt{\kappa(\Sigma^*)}}\right).}
 \label{eq:gaincost}
\end{equation}
A large formal gain therefore has a corresponding geometric cost: realizing it requires a very elongated uncertainty ellipsoid. This equality links predictive benefit and the direction-dependent precision range needed to obtain it.

For an intuitive example, take the area-preserving linear stretch $\Phi=\diag(e^{\lambda\tau},e^{-\lambda\tau})$, with prediction horizon $\tau=T-t$ and stretching rate $\lambda>0$. The matched covariance is $\rho\diag(e^{-2\lambda\tau},e^{2\lambda\tau})$. Its loss is $2\rho$ instead of the uniform-allocation loss $2\rho\cosh(2\lambda\tau)$. The instability is still present: the optimizer reduces its impact by giving the expanding direction correspondingly finer initial resolution.

\subsection{A realizable shape bound}

To represent a finite range of directional precision, impose $\kappa(Y)\le\kappa_{\max}$ in addition to Eq.~\eqref{eq:volume}. The cap describes the achievable \emph{shape} of covariance relative to $\Sigma_0$; relating it to a specific hardware or coding system additionally requires an implementation model.

For a two-dimensional area-preserving system with $Q_R=\Sigma_0=I$, the ideal covariance is feasible while $s^4\le\kappa_{\max}$. Beyond that point, the optimum uses its full permitted variance ratio, with variances $\rho\kappa_{\max}^{-1/2}$ and $\rho\kappa_{\max}^{1/2}$ along the expanding and contracting directions, respectively. Because the gain is \emph{the isotropic loss divided by the constrained minimum loss}, not the constrained loss itself, direct substitution gives
\begin{equation}
 G_{\kappa}=\begin{cases}
 \tfrac12(s^2+s^{-2}),&s^4\le\kappa_{\max},\\[3pt]
 \displaystyle\frac{s^2+s^{-2}}{s^2/\sqrt{\kappa_{\max}}+s^{-2}\sqrt{\kappa_{\max}}},&s^4>\kappa_{\max}.
 \end{cases}
 \label{eq:Gcap}
\end{equation}
In the strong-stretching limit this becomes
\begin{equation}
 \boxed{G_{\kappa}\longrightarrow\sqrt{\kappa_{\max}}.}
 \label{eq:plateau}
\end{equation}
Thus unbounded stretching need not yield unbounded \emph{realizable} improvement: once the permitted uncertainty shape is saturated, the gain approaches a finite ceiling. The absolute prediction loss can still increase rapidly; the ceiling concerns the benefit of reallocating a fixed volume relative to a uniform allocation. Equation~\eqref{eq:Gcap} makes the distinction between an ideal spectrum-based advantage and one available to a finite directional range quantitative.

A related dimension-dependent statement follows without assuming area preservation. For any positive-definite predictive metric in dimension $n$, the maximum gain under $\det Y=\rho^n$ and $\kappa(Y)\le\kappa_{\max}$ satisfies
\begin{equation}
 \boxed{1\le G_{\kappa}\le\kappa_{\max}^{(n-1)/n}.}
 \label{eq:generalbound}
\end{equation}
The upper bound is sharp in the limit that one predictive direction dominates all others (Appendix~\ref{app:bound}). In four dimensions it permits gains approaching $\kappa_{\max}^{3/4}$, so the two-dimensional square-root ceiling is not a universal rule for higher-dimensional Hamiltonian systems. The number and relative strengths of dominant stretching directions matter.

A high-resolution independent-quantization model offers one possible interpretation. When variance along direction $i$ scales as $2^{-2b_i}$, a variance ratio $\kappa_{\max}$ corresponds to a maximum directional bit-depth difference $\Delta b=\tfrac12\log_2\kappa_{\max}$ \cite{GrayNeuhoff1998}. This gives physical intuition for the shape cap while keeping the covariance model distinct from implementation-specific bit costs.

\subsection{Which directions can actually receive precision?}\label{sec:axes}

An ellipsoidal error distribution can be rotated to follow the sensitive directions only if its principal axes are accessible. Suppose instead that errors in the \emph{specified, reference-scaled state coordinates} must remain uncorrelated, so that $Y=\diag(y_1,\ldots,y_n)$ with $\prod_i y_i=\rho^n$. The covariance then has adjustable widths but cannot rotate. Direct minimization gives
\begin{equation}
 \boxed{\mathcal L_{\mathrm{diag}}^*=n\rho\left(\prod_{i=1}^n C_{ii}\right)^{1/n},\qquad
 \frac{\mathcal L_{\mathrm{diag}}^*}{\mathcal L_{\mathrm{full}}^*}
 =\left(\frac{\prod_i C_{ii}}{\det C}\right)^{1/n}\ge1.}
 \label{eq:diag}
\end{equation}
The inequality is Hadamard's determinant inequality. It gives an exact \emph{axis-accessibility penalty}: unequal diagonal sensitivities can be balanced, but correlations between accessible coordinates cannot be exploited. The general optimum and the diagonal optimum coincide if $C$ is diagonal in the accessible coordinate basis. Unlike the unrestricted gain, the diagonal constraint is tied to a physical choice of coordinates; rotating the state variables without rotating the implementation changes the feasible set.

A two-dimensional example makes the distinction visible. If $C=R_{\theta}\diag(a,a^{-1})R_{\theta}^{\mathsf T}$ with $a\ge1$, a principal axis at $\theta=\pi/4$ gives $C_{11}=C_{22}=(a+a^{-1})/2$. In that basis, $G_{\mathrm{diag}}=1$ even though $G_{\mathrm{full}}=(a+a^{-1})/2$. Thus a device restricted to independent per-coordinate precision may gain nothing precisely when the predictive geometry calls for a rotated allocation.

For brevity, denote the anisotropy limit $\kappa_{\max}$ by $K$ in the following constrained-allocation results. With a two-dimensional condition-number cap $\kappa(Y)\le K$, a coordinate-diagonal allocation $Y=\rho\diag(y,y^{-1})$ has the closed-form solution
\begin{equation}
 y_{\mathrm{diag}}^*=\operatorname{clip}\left(\sqrt{C_{22}/C_{11}},K^{-1/2},K^{1/2}\right),\quad
 G_{\mathrm{diag},K}=\frac{\tr C}{C_{11}y_{\mathrm{diag}}^*+C_{22}/y_{\mathrm{diag}}^*}.
 \label{eq:diagcap}
\end{equation}
Here $\operatorname{clip}(x,l,u)=\min\{u,\max\{l,x\}\}$. Equations~\eqref{eq:diag}--\eqref{eq:diagcap} permit a direct comparison of unrestricted directions and physically fixed coordinate axes at the same volume and condition-number budget.

\subsection{When strong stretching suppresses differences between tasks}\label{sec:erasure}

Directional range and accessible axes govern the gain attainable at a given horizon. A further consequence of the same anisotropy cap concerns \emph{task specificity}: when one stretching direction dominates, distinct terminal requirements can favor nearly the same allocation. Let $B_T=\Phi_{t,T}\Sigma_0^{1/2}$, and consider two fixed positive-definite requirements $Q_1,Q_2$. If $\sigma_2(B_T)/\sigma_1(B_T)\to0$ as the horizon increases, then at any fixed finite cap $K>1$ the two determinant-normalized optimal shapes converge to the same moving shape:
\begin{equation}
 \boxed{\frac{Y_{Q_i,K}^*(T)}{\rho}-\Big[\alpha_K v_Tv_T^{\mathsf T}+\beta_K(I-v_Tv_T^{\mathsf T})\Big]\longrightarrow0,\quad i=1,2,}
 \label{eq:erasure}
\end{equation}
where the limit is in matrix norm, $v_T$ is a leading right singular vector of $B_T$, $\alpha_K=K^{-(n-1)/n}$, and $\beta_K=K^{1/n}$. The direction $v_T$ need not itself converge. The corresponding loss penalty from using the other task's capped optimal shape tends to one (Appendix~\ref{app:erasure}).

The physical interpretation is a competition between \emph{task choice} and \emph{dynamical stretching}. A task can distinguish different directions at finite horizon, but once one direction dominates and the allowed anisotropy saturates, all positive-definite terminal tasks favor placing the smallest permitted variance along that direction. With multiple comparable leading directions, task dependence can persist. The limit describes capped, tangent-model allocations; Section~\ref{sec:finite} separately evaluates finite-amplitude prediction. Earlier singular-vector studies have analyzed long-horizon changes of norm sensitivity \cite{ReynoldsErrico1999,Kuang2004}; Eq.~\eqref{eq:erasure} makes the associated consequence for \emph{fixed-volume, capped covariance allocation} explicit.

\section{Examples in Hamiltonian dynamics}\label{sec:examples}

\subsection{Standard map: strongly heterogeneous benefit across initial states}

Consider the area-preserving standard map \cite{Chirikov1979}, with angular coordinate $\theta$, conjugate momentum $p$, and nonlinearity parameter $K_s$:
\begin{align}
 p_{j+1}&=p_j+K_s\sin\theta_j\pmod{2\pi},\\
 \theta_{j+1}&=\theta_j+p_{j+1}\pmod{2\pi}.
 \label{eq:standard}
\end{align}
For the coordinate ordering $(\theta,p)$, its one-step Jacobian is
\begin{equation}
 J_j=\begin{pmatrix}1+K_s\cos\theta_j&1\\K_s\cos\theta_j&1\end{pmatrix},\qquad \det J_j=1.
 \label{eq:jacstandard}
\end{equation}
Multiplying $N$ successive Jacobians produces the finite-horizon tangent map $\Phi_N$. We use $Q_R=\Sigma_0=I$, $K_s=1.2$, and the formulas in Eqs.~\eqref{eq:Gandk} and \eqref{eq:Gcap}.

\begin{figure}[t]
\centering
\includegraphics[width=0.98\linewidth]{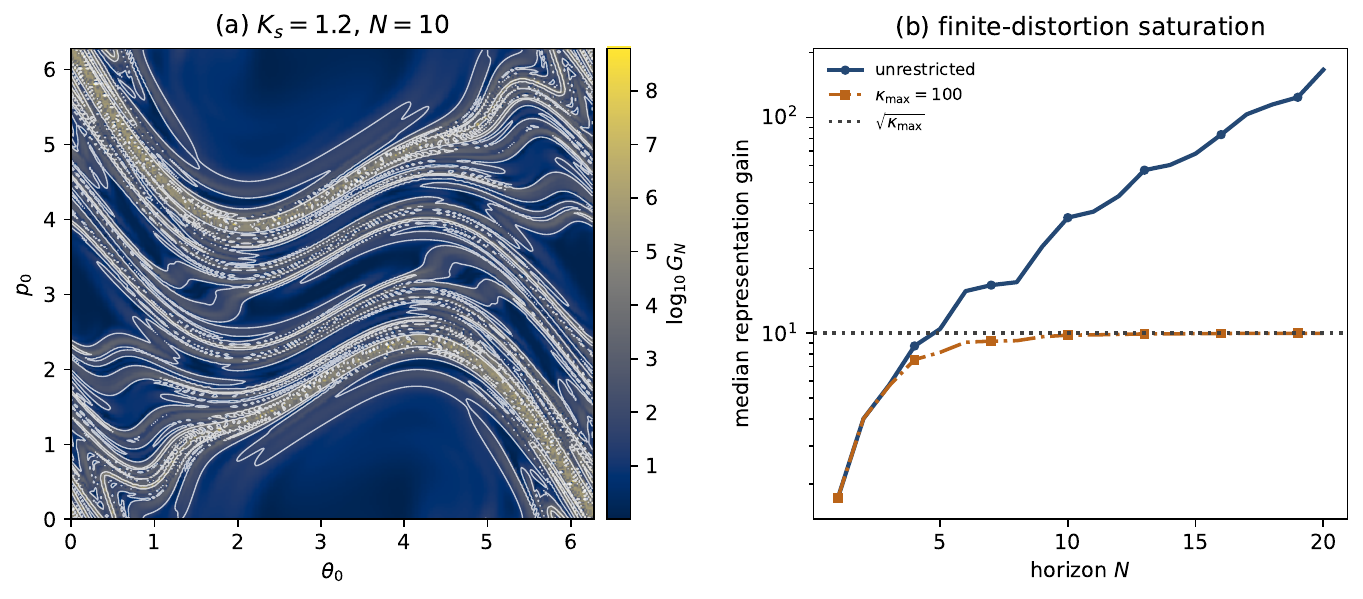}
\caption{Standard-map gain for $K_s=1.2$. (a) The spatial distribution of $\log_{10}G$ over the $(\theta_0,p_0)$ torus at $N=10$; luminance and overlaid contours remain readable in grayscale. (b) Median ideal gain (solid line with circular markers), capped gain for $\kappa_{\max}=100$ (dash--dot line with square markers), and the theoretical ceiling $\sqrt{\kappa_{\max}}=10$ (dotted horizontal line). The panels evaluate the finite-time tangent model.}
\label{fig:standard}
\end{figure}

Figure~\ref{fig:standard}a maps the resulting ideal gain over initial conditions at $N=10$. Its filamentary spatial variation comes from nonuniform finite-time stretching: some initial states strongly reward directional allocation, whereas nearby states offer far less improvement. Figure~\ref{fig:standard}b compares the median ideal and capped gains as the horizon increases. For 20,000 uniformly sampled initial states (seed 12345), the median ideal gain rises from $1.723$ at $N=1$ to $166.566$ at $N=20$. With $\kappa_{\max}=100$, the corresponding medians are $1.723$ and $9.991$, approaching the predicted ceiling of 10.

\begin{figure}[t]
\centering
\includegraphics[width=0.99\linewidth]{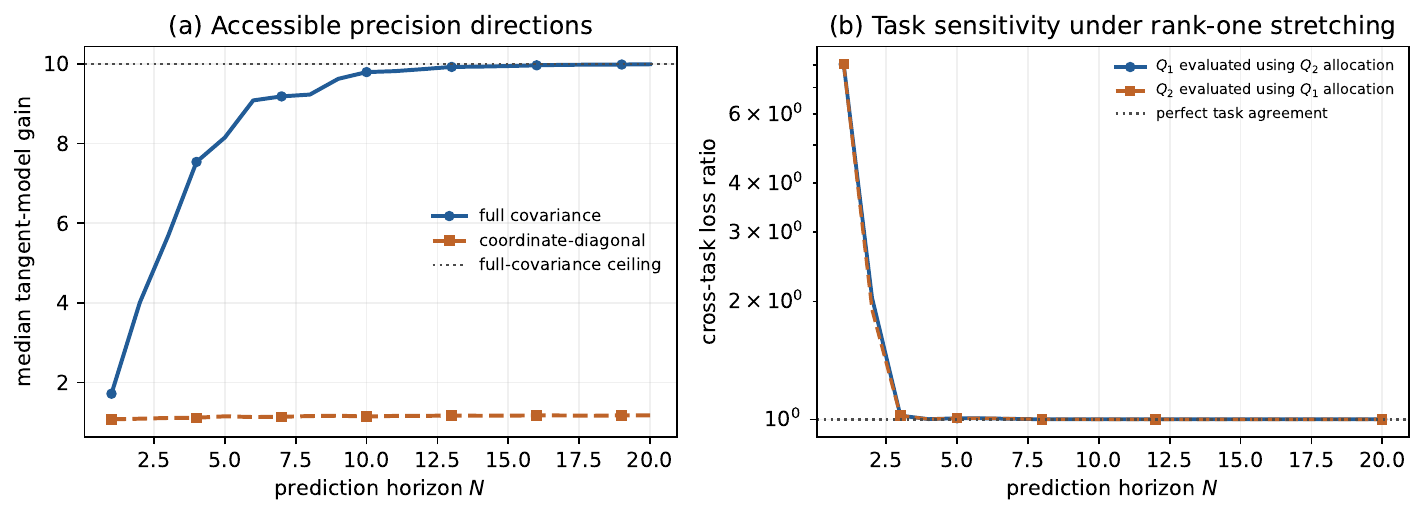}
\caption{Two consequences of finite directional control in the standard map. (a) Median tangent-model gains for unrestricted covariance orientation (solid circles), covariance diagonal in the displayed $(\theta,p)$ coordinates (dashed squares), and the unrestricted-orientation ceiling (dotted), all with $\kappa_{\max}=100$; the same 20,000 initial states are used. (b) Cross-task loss ratios for one specified trajectory and two terminal metrics, using capped task-optimal covariances; unity denotes no mismatch penalty. Colour is redundant with line styles and markers.}
\label{fig:access}
\end{figure}

The ability to \emph{rotate} the allocation matters as much as its available elongation. Figure~\ref{fig:access}a applies Eq.~\eqref{eq:diagcap} to the same 20,000 initial states and the same cap $K=100$. At $N=20$, the median gain is $9.991$ when covariance axes are free to rotate but only $1.180$ when the covariance must be diagonal in $(\theta,p)$. The physical gap is not additional stretching; it comes from the alignment of the sensitive direction with the available precision axes.

To illustrate Eq.~\eqref{eq:erasure}, Fig.~\ref{fig:access}b uses one strongly stretched standard-map trajectory, $x_0=(4.76047,2.05419)$, and two terminal requirements $Q_1=\diag(4,1/4)$ and $Q_2=\diag(1/4,4)$. At $K=100$, using the $Q_2$-optimal covariance for the $Q_1$ task costs a factor $8.031$ at $N=1$, but about $1.022$ at $N=3$ and $1.006$ at $N=5$. The changing cross-task penalty follows the finite-time separation of sensitive directions along this trajectory.

\subsection{Double pendulum: gain and task dependence in four dimensions}

The planar double pendulum gives a continuous-time mechanical example with canonical coordinates $\bm{x}=(q_1,q_2,p_1,p_2)$, where $q_i$ are angles and $p_i$ their conjugate momenta. Using unit masses, unit lengths, and unit gravitational acceleration, and defining $h=\cos(q_1-q_2)$, its Hamiltonian is \cite{Shinbrot1992,StachowiakOkada2006}
\begin{equation}
 H(\bm{x})=\frac{p_1^2-2hp_1p_2+2p_2^2}{2(2-h^2)}-2\cos q_1-\cos q_2.
 \label{eq:Hdp}
\end{equation}
The orbit follows Hamilton's equations. Its tangent map satisfies $\dot\Phi=J_4\nabla^2H(\bm{x}(t))\Phi$ with $\Phi(0)=I_4$ and the canonical symplectic matrix $J_4=\bigl(\begin{smallmatrix}0&I_2\\-I_2&0\end{smallmatrix}\bigr)$. Consequently, the exact tangent map is symplectic and volume preserving. The numerical calculation uses the same nondimensional reference scaling for all four canonical coordinates; comparisons with other physical unit conventions require transforming $\Sigma_0$ and $Q_R$ consistently.

\begin{figure}[t]
\centering
\includegraphics[width=0.77\linewidth]{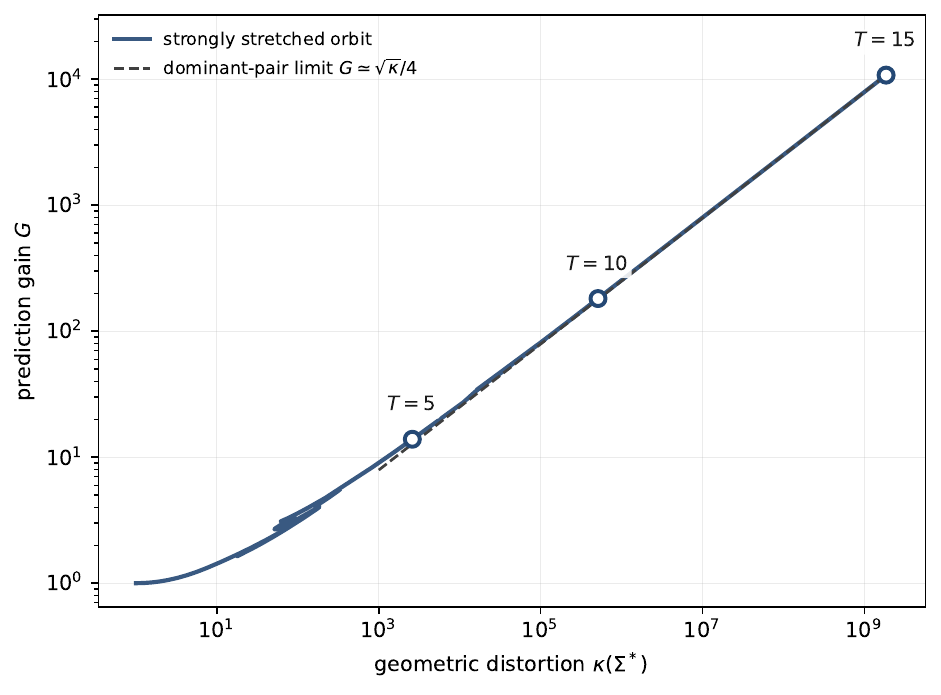}
\caption{Four-dimensional double-pendulum gain versus ideal covariance condition number for the strongly stretched initial state. The solid curve is the tangent calculation; open circles locate the labelled prediction times. The dashed curve is $G\simeq\sqrt{\kappa}/4$, the asymptote for one dominant reciprocal singular-value pair. Line style and markers distinguish the curves without relying on colour.}
\label{fig:double}
\end{figure}

We contrast initial states $(0.2,0.1,0,0)$ (energy $H\simeq-2.95514$) and $(2,1,0,0)$ ($H\simeq0.29199$). Both use $Q_R=\Sigma_0=I$. The former remains weakly stretched over the inspected interval; at $T=15$ its ideal gain is about $1.47$. The latter develops much stronger finite-time deformation: its gains are approximately $13.88$, $181.53$, and $1.076\times10^4$ at $T=5$, 10, and 15. Figure~\ref{fig:double} traces the gain against required covariance anisotropy. In four-dimensional symplectic dynamics, if one singular-value pair $(s,s^{-1})$ dominates while the other remains of order one, $\tr C\sim s^2$ and $\det C=1$. Equation~\eqref{eq:gain} then gives $G\sim s^2/4\sim\sqrt{\kappa}/4$, explaining the dashed asymptote and its coefficient of one quarter. The relevant diagnostic is the measured stretching spectrum over each finite prediction horizon.

The task itself also matters. At the higher-energy initial state with horizon $T=0.5$, define two equal-determinant terminal metrics
\begin{align}
 Q_R^{(1)}&=\diag(4,4^{-1/3},4^{-1/3},4^{-1/3}),\\
 Q_R^{(2)}&=\diag(4^{-1/3},4,4^{-1/3},4^{-1/3}).
 \label{eq:twoQ}
\end{align}
The first emphasizes $q_1$, whereas the second emphasizes $q_2$. Their leading optimal-precision eigenvectors have absolute inner product $0.283$ (an acute angle of about $74^\circ$ in the chosen normalized coordinates). Thus \emph{even for the same orbit and horizon}, changing the terminal observable changes the direction that deserves the finest present resolution. This is why stretching alone is insufficient to answer the allocation question.

\subsection{A direct finite-amplitude check}\label{sec:finite}

\begin{figure}[t]
\centering
\includegraphics[width=0.78\linewidth]{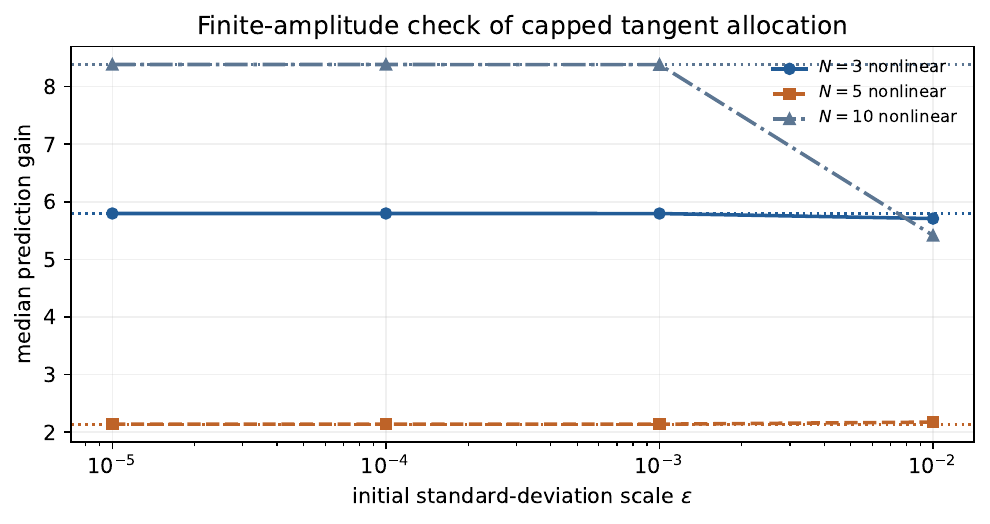}
\caption{Finite-amplitude check in the standard map. Solid, dashed, and dash--dot lines with distinct markers show median empirical gains at horizons $N=3,5,10$, respectively, from 24 reference states and 8,192 perturbations per state and condition. Horizontal dotted lines in matching grey or colour shades show the median tangent-model predictions. The common covariance determinant is $\varepsilon^4$, and $\kappa_{\max}=100$. The experiment assesses the finite-amplitude validity of the tangent-optimal capped allocation.}
\label{fig:finite}
\end{figure}

The preceding figures use tangent maps to predict the response to local errors. We next propagate finite Gaussian perturbations through the \emph{nonlinear} standard map to test the range of that prediction. Write the initial covariance as $\Sigma=\varepsilon^2 S$ with $\det S=1$, so that $\varepsilon$ is the standard deviation of the reference-isotropic allocation. We compare isotropic and tangent-optimal capped covariances at the same determinant $\varepsilon^4$, with $\kappa_{\max}=100$. For each of 24 independently sampled reference states, 8,192 antithetically paired Gaussian perturbations are propagated to $N=3,5,10$. Terminal angular and momentum differences are taken on the torus, and gain is the ratio of the two empirical mean-squared errors. Figure~\ref{fig:finite} shows the median of \emph{per-trajectory} gains as $\varepsilon$ varies; dotted levels denote the corresponding tangent predictions. At $N=10$ the median gain is $8.384$ at $\varepsilon=10^{-5}$, close to the tangent prediction $8.383$, but decreases to $5.416$ at $\varepsilon=10^{-2}$. Thus the allocation benchmark is accurate for sufficiently localized perturbations, while finite-amplitude deformation can weaken its benefit.

\section{Interpretation and range of applicability}\label{sec:meaning}

The central result is a predictive allocation benchmark with \emph{two distinct limits to realizing it}. The terminal requirement and trajectory define the present sensitivity metric [Eq.~\eqref{eq:A}], which determines the ideal fixed-volume covariance [Eq.~\eqref{eq:sigmaopt}]. Its benefit depends on how much anisotropy can be realized [Eqs.~\eqref{eq:Gcap} and \eqref{eq:generalbound}] and on whether those principal directions can be accessed [Eq.~\eqref{eq:diag}]. Figure~\ref{fig:access} shows that the latter restriction is not a minor technical detail: in a rotated sensitive geometry, an otherwise adequate direction-dependent precision range may provide little gain in a coordinate-fixed implementation.

The relation to existing methods is specific. Singular-vector approaches characterize the growth and task dependence of individual perturbations \cite{Palmer1998}; optimal-design and continuous allocation methods supply determinant--trace tools \cite{Pukelsheim2006,GrayNeuhoff1998}; and nonlinear optimal perturbations address finite-amplitude effects under different objectives \cite{Mu2003}. Here the optimization variable is the \emph{whole local covariance shape}, with the same uncertainty volume in every comparison and an explicitly selected terminal loss. The gain--anisotropy equations say how much the geometry can help and how much representational distortion the ideal improvement requires. Assessing realized forecasts further requires an explicit account of encoding, noise, and implementation \cite{Cooper2020}.

The direct nonlinear propagation in Fig.~\ref{fig:finite} quantifies the perturbation-scale range over which the tangent allocation accurately predicts the empirical gain for the sampled standard-map trajectories. Appendix~\ref{app:nonlinear} supplies a complementary exact example: the finite-amplitude optimum can change while the tangent metric stays fixed. The pair of examples distinguishes the tangent allocation benchmark from the finite-amplitude covariance optimum. The closed-form results assume positive-definite terminal requirements; a task selecting only part of the state requires an explicit allocation rule for directions outside its sensitivity subspace.

Related preprints by the author examine how the \emph{amount} of required precision changes with forecast horizon and target accuracy (S.-J. Kim, arXiv:2609.38674 and arXiv:2610.00352, 2026). The present result instead determines how a specified, fixed local uncertainty volume is distributed among directions for one terminal prediction task. This separates geometrical allocation from resource-scaling laws.

\section{Conclusion}\label{sec:conclusion}

Finite-time prediction defines a geometry of present-state sensitivity relative to a chosen terminal task. At fixed local uncertainty volume, the inverse predictive metric uniquely balances the task-weighted terminal error, and its spectrum determines the gain over uniform allocation. The attainable gain is shaped by two distinct capabilities: how much directional precision contrast can be realized and whether the covariance axes can follow the sensitive directions. In the limit of a single dominant stretching direction, a fixed anisotropy cap also drives allocations for different terminal tasks toward a common shape. The standard map exposes the impact of these restrictions, the double pendulum illustrates task-dependent geometry in four dimensions, and finite-amplitude propagation identifies where the local prediction remains accurate. These results provide a quantitative basis for adaptive sensing, precision-aware state encoding, and numerical forecasting. A natural next step is to connect the covariance-level bounds to integer-bit and sensor-noise constraints and to evaluate task-matched allocations in online prediction workflows.

\section*{Acknowledgements}
This work was supported by the research grant SP014 from SOBIN Institute, LLC.

\section*{Statements and Declarations}
\noindent\textbf{Funding.} No external funding was received for this work.

\noindent\textbf{Competing interests.} The author declares no competing interests.

\noindent\textbf{Data availability.} The numerical data and code supporting the findings of this study are available from the corresponding author upon reasonable request.

\noindent\textbf{Use of AI tools.} OpenAI ChatGPT assisted with language editing, manuscript organization, research-code preparation, and review of mathematical and numerical presentation. The author retains responsibility for verification of scientific claims, calculations, references, and the submitted manuscript.

\appendix
\section{Coordinate changes preserve the allocation result}\label{app:coordinates}
For an invertible coordinate transformation $\bm y=S\bm x$, consistent tensors transform as $\Sigma_y=S\Sigma_xS^{\mathsf T}$, $\Sigma_{0,y}=S\Sigma_{0,x}S^{\mathsf T}$, and $A_y=S^{-\mathsf T}A_xS^{-1}$. Hence $\tr(A_y\Sigma_y)=\tr(A_x\Sigma_x)$ and $\det(\Sigma_{0,y}^{-1}\Sigma_y)=\det(\Sigma_{0,x}^{-1}\Sigma_x)$. The eigenvalues of $\Sigma_0A$ (and therefore the ratio in Eq.~\eqref{eq:gain}) are unchanged, since $(\Sigma_0 A)_y=S(\Sigma_0A)_xS^{-1}$.

\section{A general upper bound for bounded anisotropy}\label{app:bound}
Write $Y=\rho\overline Y$ with $\det\overline Y=1$ and $\kappa(\overline Y)\le\kappa_{\max}$. If $y_{\min}$ is the smallest eigenvalue of $\overline Y$, then every other eigenvalue is at most $\kappa_{\max}y_{\min}$. Consequently,
\begin{equation}
 1=\det\overline Y\le y_{\min}^n\kappa_{\max}^{n-1},\qquad
 y_{\min}\ge\kappa_{\max}^{-(n-1)/n}.
\end{equation}
For any $C>0$, $\tr(CY)\ge\rho y_{\min}\tr C$, giving Eq.~\eqref{eq:generalbound}. Sharpness follows by choosing $\overline Y$ with eigenvalues $(\kappa_{\max}^{-(n-1)/n},\kappa_{\max}^{1/n},\ldots,\kappa_{\max}^{1/n})$ and letting one eigenvalue of $C$ dominate, while the others remain positive.

\section{Derivations for axis restrictions and task erasure}\label{app:erasure}

For a coordinate-diagonal $Y=\diag(y_i)$, $\tr(CY)=\sum_i C_{ii}y_i$. Holding $\prod_i y_i=\rho^n$ fixed, the arithmetic--geometric mean inequality gives
\begin{equation}
 \sum_i C_{ii}y_i\ge n\rho\left(\prod_i C_{ii}\right)^{1/n},
 \qquad y_i^*=\frac{\rho(\prod_j C_{jj})^{1/n}}{C_{ii}}.
\end{equation}
The equality condition, Proposition~\ref{prop:opt}, and Hadamard's inequality establish Eq.~\eqref{eq:diag}. Minimizing $C_{11}y+C_{22}/y$ on $[K^{-1/2},K^{1/2}]$ gives Eq.~\eqref{eq:diagcap}.

For Eq.~\eqref{eq:erasure}, write $C_i(T)=B_T^{\mathsf T}Q_iB_T$. If $\sigma_2(B_T)/\sigma_1(B_T)\to0$ and $Q_i$ is fixed and positive definite, then $C_i(T)/\tr C_i(T)-v_Tv_T^{\mathsf T}\to0$ in operator norm. The compact feasible set of determinant-one positive-definite $S$ with $\kappa(S)\le K$ has a unique minimizer of $v_T^{\mathsf T}Sv_T$, namely $S=\alpha_Kv_Tv_T^{\mathsf T}+\beta_K(I-v_Tv_T^{\mathsf T})$. Indeed, $\lambda_{\min}(S)\ge\alpha_K$, and equality with determinant and ratio cap fixes all other eigenvalues to $\beta_K$. Uniform convergence of the normalized objectives on this compact set, together with uniqueness of the limiting minimizer, proves that both task-optimal shapes approach this same moving matrix. The associated cross-task loss ratio approaches one because the denominator stays bounded away from zero after normalized-trace scaling.

\section{An exact nonlinear example defining the local boundary}\label{app:nonlinear}
Consider the smooth area-preserving map $F_c(x,y)=(\lambda x,\lambda^{-1}y+cx^2)$, with $\lambda>0$. Its tangent map at the origin is $\diag(\lambda,\lambda^{-1})$, independent of the nonlinear coefficient $c$. For a centered Gaussian initial perturbation with covariance
\begin{equation}
 \Sigma=\varepsilon^2\begin{pmatrix}s&k\\k&(1+k^2)/s\end{pmatrix},\qquad s>0,
\end{equation}
we have $\det\Sigma=\varepsilon^4$. The exact mean-square terminal error (with $Q_R=I$) is
\begin{equation}
 \mathcal L_c(s,k)=\varepsilon^2\left[\lambda^2s+\frac{\lambda^{-2}(1+k^2)}{s}\right]+3c^2\varepsilon^4s^2.
 \label{eq:shear}
\end{equation}
The optimum occurs at $k=0$. For $c=0$ it has $s=\lambda^{-2}$, whereas for any $c\ne0$ and $\varepsilon>0$, the derivative of Eq.~\eqref{eq:shear} is positive at $s=\lambda^{-2}$, so the optimal $s$ shifts. Thus two area-preserving maps with the same tangent metric can have different finite-amplitude covariance optima. Conversely, a nonlinear twist $F(z)=R_{g(\|Lz\|^2)}Lz$, with $L=\diag(\lambda,\lambda^{-1})$ and $R$ a rotation, preserves $\|F(z)\|^2=\|Lz\|^2$ at every amplitude; there the local covariance optimum is exact at finite amplitude too. These examples identify the role of the terminal loss geometry rather than merely the presence or absence of nonlinear terms.

\section{Numerical reproduction notes}\label{app:numerics}
For the standard map, the $280\times280$ phase-space grid was propagated using products of the exact one-step Jacobian in Eq.~\eqref{eq:jacstandard}. The median curves, including the coordinate-diagonal comparison in Fig.~\ref{fig:access}a, use the same 20,000 independent uniformly selected initial states (seed 12345). Figure~\ref{fig:access}b uses the specified single trajectory and endpoint metrics; its per-task capped optima are obtained by diagonalizing the two weighted pullback matrices. For Fig.~\ref{fig:finite}, 24 reference states (seed 20261118) and 4,096 Sobol Gaussian points plus their antithetic partners (seed 314159) yield 8,192 perturbations per state and condition. The physical nonlinear map is iterated, with differences measured on the $2\pi$-periodic torus. The underlying code and per-trajectory losses are available from the corresponding author upon reasonable request, enabling the aggregate curves to be independently checked. For the double pendulum, the Hamiltonian and the $4\times4$ variational matrix were integrated together using adaptive DOP853 with relative tolerance $10^{-10}$ and absolute tolerance $10^{-12}$. The Hamiltonian vector-field Jacobian was computed using complex-step differentiation. For the two reported orbits, relative energy drift remained below $7.4\times10^{-10}$ and $|\det\Phi-1|$ below $3.5\times10^{-10}$ through $T=15$. The reproducibility scripts record numerical diagnostics and regenerate all five figures: the conceptual illustration, the two Hamiltonian-system examples, and the allocation and finite-amplitude comparisons.

\end{document}